# Epidemiological and public health requirements for COVID-19 contact tracing apps and their evaluation

Vittoria Colizza[1,2], Eva Grill[3], Rafael Mikolajczyk[4], Ciro Cattuto[5,6], Adam Kucharski[7], Steven Riley[8], Michelle Kendall[9,10], Katrina Lythgoe[9], Lucie Abeler-Dörner[9], Chris Wymant[9], David Bonsall[9], Luca Ferretti[9,*], Christophe Fraser[9,*]

[1] INSERM, Sorbonne Université, Institut Pierre Louis d'Epidémiologie et de Santé Publique, IPLESP, Paris, France

[2] Tokyo Tech World Research Hub Initiative, Institute of Innovative Research, Tokyo Institute of Technology, Tokyo, Japan

[3] Institute für Medical Information Processing, Biometry and Epidemiology, Ludwig-Maximilians University Muenchen, Muenchen, Germany

[4] Institute for Medical Epidemiology, Biometrics and Informatics, Interdisciplinary Center for Health Sciences, Martin Luther University Halle-Wittenberg, Halle, Germany

[5] University of Turin, Turin, Italy

[6] ISI Foundation, Turin, Italy

[7] Centre for Mathematical Modelling of Infectious Diseases, London School of Hygiene & Tropical Medicine, London, UK

[8] MRC Centre for Global Infectious Disease Analysis, School of Public Health, Imperial College London, UK

[9] Big Data Institute, Li Ka Shing Centre for Health Information and Discovery, Nuffield Department of Medicine, University of Oxford, UK

[10] Department of Statistics, University of Warwick, Warwick, UK

* Corresponding authors. Email: luca.ferretti@bdi.ox.ac.uk, christophe.fraser@bdi.ox.ac.uk

**Digital contact tracing is a public health intervention. It should be integrated with local health policy, provide rapid and accurate notifications to exposed individuals, and encourage high app uptake and adherence to quarantine. Real-time monitoring and evaluation of effectiveness of app-based contact tracing is key for improvement and public trust.**

In early 2020, COVID-19 caused a world-wide pandemic despite the fact that many countries had epidemic preparedness plans. Experienced teams of contact tracers were in place to interrupt the early phases of domestic transmission, but were soon overwhelmed in many places by the number of cases. In the absence of a vaccine, governments looked at additional non-pharmaceutical interventions to prevent the spread of infections. As SARS-CoV-2 is an airborne respiratory pathogen, physical separation of individuals through generalised lockdowns and travel restrictions proved effective, but came at great social, psychological and economic cost. Most countries also increased their focus on case-driven interventions such as identification of clusters and cases, and tracing their contacts. Lockdowns are essentially wholly non-specific quarantines which affect the whole population. Contact tracing, on the other hand, restricts quarantining to those with recent contact with known or suspected cases. A combination of the latter with less disruptive preventive measures like physical distancing and the wearing of face masks is clearly the preferred way to suppress the epidemic in the absence of a vaccine.

Contact tracing is a standard tool for outbreak control of infectious diseases, and has been used at least since John Snow investigated a cholera outbreak in London in 1854. Contact tracing traditionally involves the analysis of social contact information acquired by interview with confirmed or suspected cases. Contacts are given specific health advice, according to the characteristics of the pathogen and social context of transmission. Modern day partner notification in sexual health, for example, is a form of contact tracing that delivers treatments to sexual partners of cases that may have occurred months or years in the past. Before the COVID-19 pandemic, manual contact tracing programmes were established in many countries, but were not equipped or staffed to the extent required to prevent a pandemic caused by a highly infectious virus like SARS-CoV-2. In addition, contact tracing for COVID-19 proved especially challenging, owing to the short generation time of SARS-CoV-2 and the high rates of pre-symptomatic transmission. The most successful programmes in Singapore and South Korea focused on speed and effectiveness and quickly linked tracing to high throughput testing. Several modelling studies of the COVID-19 epidemic identified that contact tracing needs to reach contacts within 48 hours from the onset of symptoms in the index case to stop transmission chains and prevent onward infections [Kretzschmar 2020, Kucharski 2020].

Digital contact tracing apps have been proposed to help control the spread of COVID-19 [Ferretti 2020] and now represent a key component of many national strategies for suppressing the epidemic [Hinch 2020, O'Neill 2020], alongside bolstered conventional manual contact tracing programmes. These apps implement privacy-preserving proximity detection and exposure notification.

App-based contact tracing is first and foremost a public health intervention. Designing an effective app requires expertise from diverse fields including engineering [Sattler 2020], information security [Troncoso 2020, PEPP-PT, Vaudenay 2020a,b], ethics [Parker 2020, Morley 2020], and behavioural and social sciences [Abeler 2020, Farronato 2020]. Contributions from all these areas are essential yet outside the scope of this commentary. We will therefore only touch on them where they intersect with epidemiological and public health considerations.

SARS-CoV-2 is likely to become endemic in many parts of the world, and while widespread immunity from vaccination is the final goal, there is still no certainty about how quickly vaccination will become available across countries and age groups and how long its protection will last. For the foreseeable future, most countries will continue to rely on a combination of different measures, including vaccination, social distancing, mask wearing and contact tracing. For contact tracing apps to achieve their primary purpose of substantially reducing COVID-19 transmission, epidemiological considerations must be at the heart of their design.

We present five key epidemiological and public health requirements which COVID-19 contact tracing apps should satisfy.

1. **Integration with local health policy**
   The advice given by an app notification should be adjustable to remain consistent with local health policies. Ideally the app should be integrated within the full range of public health interventions, such as providing information on the steps to follow upon notification, access to testing, medical care and advice on isolation, and work in conjunction with conventional contact tracing where available. Whenever possible, the app-based and conventional contact tracing processes should be integrated, especially downstream of exposure notifications, when app users are asked to contact the relevant health authority (e.g., via dedicated call centers). Local health authorities need to scale up resources for local testing and contact tracing to handle the additional caseload.

   Information about exposure to a case that is provided via an app should have the same implications as such information provided through other routes. This means for example that it can initiate quarantine, and provide recognised justification for absence from work, school, and associated financial and logistical supports. Otherwise, a large proportion of the population is unlikely to quarantine as recommended by the app, and its effect will be restricted. If this equality is not legally granted, some people may also prefer not to use the app at all, in order to avoid receiving a notification that they are unable to adhere to.

   In order to allow resumption of travel without increasing the risk of epidemic resurgence, apps should be **interoperable** across countries [EC 2020]. Without this, users would have to install multiple apps, with a likely detrimental effect on accuracy, uptake, and adherence. Interoperability should also extend to the public health aspects: the app's integration into local health policy should adapt to the country the user is in, for example with regards to isolation and further advice.

2. **High user uptake and compliance**
   Even at low levels of **uptake**, apps can reduce transmission and can have a protective effect on the population, including for those without smartphones [Hinch 2020a]. However, the effectiveness of app-based contact tracing increases with the number of users [Ferretti 2020, Lambert 2020, Cencetti 2020, Moreno Lopez 2020]. An effective communication strategy explaining the app functioning and confidentiality policy is essential to increase adoption [Montagni 2020]. Once installed, an app will only affect onwards transmission if users follow the recommendations it issues. Trust in the app and a positive user experience are therefore essential components for digital contact tracing to be effective. Any design choices which could hinder **adherence** should be avoided. The design, implementation and deployment of the app should also be guided by equality considerations, making it accessible for harder-to-reach populations such as older adults, persons from deprived regions, and migrant populations. Deployment strategies should involve local communities and encourage high local rates of adoption, increasing both the perceived individual benefit of the app and its local effectiveness [Farronato 2020].

3. **Quarantine infectious individuals as accurately as possible**
   The purpose of contact tracing apps is to quarantine individuals who are or will become infectious. Two things should therefore be minimised: failure to quarantine infectious individuals, and the time spent in quarantine by non-infectious individuals.

   To achieve optimal performance, the algorithm implemented by the app must be **tunable** and should be tuned as new data on transmission risk becomes available. In a rapidly developing epidemic, our knowledge of the disease will be continuously improving. It may vary across populations and social networks, and over time through the impact of interventions such as physical distancing, wearing of masks, or protocols implemented in specific settings. Tuning the required distance and time of exposure to send out notifications results in varying sensitivity and specificity of the app, critically affecting its ability to reduce transmission and the portion of the population that is quarantined [Cencetti 2020, Briers 2020, Wilson 2020]. To freeze an algorithm before the app is released, with no capacity for change, reduces the expected performance of the app.

   Furthermore, it should be possible to send out **early release notifications**. These may be needed as a means of correction if a quarantine notification is sent out erroneously or needs to be updated. This could occur due to a malfunction or through malicious use, or more simply (and perhaps frequently) in apps in which an index case can trigger an alert based on reporting symptoms, which may later be deemed not infected by a negative test result [Hinch 2020].

4. **Rapid notification**
   The time between onset of symptoms in an index case and the quarantine of their contacts is of key importance to COVID-19 contact tracing; any delay reduces its effectiveness [Ferretti 2020, Kretzschmar 2020]. Where a design feature introduces a delay, such as awaiting confirmatory test results, it should only be implemented if the delay is outweighed by other gains such as in specificity, uptake, adherence, etc. The relative impacts of these factors should be quantitatively compared at the design stage in open-source models [Hinch 2020b, Gupta 2020]. If the delays exceed the period in which most contacts transmit the disease, the app will fail to have an impact on reducing transmission.

5. **Ability to evaluate effectiveness transparently**
   The public must be provided with evidence that notifications are based on the best available data. The contact tracing algorithm should therefore be transparent, auditable, under oversight, and subject to review.

   Aggregated data (not linked to individuals) is essential for evaluating and improving the performance of the app. Although some of this information could perhaps be gained via surveys, there are strong practical and ethical justifications for gathering these data via the app itself. These justifications are particularly concerned with the speed and scale of the epidemic, and the huge social and economic costs of failing to control it. Aggregated statistics should also be used for public communication, to provide feedback to users and the public in order to increase trust and engagement. A good example is provided by the dashboards for the Italian [Immuni] and the Swiss [SwissCovid] app.

   a. **Real-time summary statistics**
      Aggregated summary statistics such as the numbers of adopters, index cases and contacts notified by the app should be available. This data is crucial for evaluating the effects of the app and rapidly identifying malfunctions or malicious use, as well as being extremely valuable for public health planning.
   b. **Geographical summary statistics**
      Knowledge of local uptake – at a sufficiently coarse-grained spatial resolution – is key for assessing the app's effectiveness and the reliability of its evaluation of individual risk. For example, individuals in areas with low app uptake and high incidence of COVID-19 could erroneously be given the impression that they are at low risk, especially in contexts where lower app uptake is correlated with greater transmission. As apps in Europe do not geolocate individuals, this additional information can be provided by the user once activating the app, if in compliance with privacy regulations of the country, or through surveys.
   c. **Transparency of public health response around digital contact tracing**
      App-based contact tracing needs to dovetail with a complex response system that is only partially digital, and places additional demands on different capabilities such as logistics of testing, organization of call centers for notified users, manual contact tracing, follow-up of quarantined individuals, etc. Often, these capabilities are provided by institutions and systems whose data are only loosely – if at all – integrated, e.g., because of the inherent regional segmentation of many national healthcare systems. Improved transparency and integration of the non-digital processes around app-based contact tracing is a crucial precondition to evaluating their effectiveness.

Most countries which have implemented app-based contact tracing as part of their COVID-19 response have now reached the point where they need to evaluate such apps as effective public health tools. This is far from easy, especially as the evaluation needs to take into account the effectiveness of other components of the overall response. We recommend that multiple independent approaches should be used in these evaluations and metrics of success and failure should be decided upon in advance.

Real-time monitoring should be performed whenever possible using aggregated data automatically retrieved from active apps. So far, a proof of principle evaluation is available for the Swiss app [Salathé 2020].

Given the reduced amount of data potentially available due to privacy-preserving designs, a more detailed assessment should be performed in conjunction with traditional contact tracing. For example, index cases seeking healthcare could be asked if they routinely use the app; if so, contacts identified by traditional contact tracing could be asked whether they used the app and received a digital notification. Follow-up assessments of the positivity rate of tests of notified individuals can then be performed. Ethical and privacy considerations need to be assessed in each country.

Effectiveness of app-based contact tracing should also be evaluated from independent observational studies, such as cross-sectional or longitudinal surveys, epidemiological analyses [Kendall 2020] and experimental studies [https://www.lamoncloa.gob.es/lang/en/gobierno/news/Paginas/2020/20200803radarcovid.aspx]. Although contact tracing apps are not always included in current standard frameworks for digital health technologies [NICE 2019], their evaluation should follow these frameworks as closely as possible.

Appropriate studies should evaluate both effectiveness and the process of implementation. Process evaluation is key to understand potential reasons for lack of effectiveness that may not be attributable to functionality of the app itself but e.g. to poor uptake or adherence, inadequate handling of cases by the health system or insufficient information. Likewise, any potential harms should be rigorously considered as part of a systematic risk assessment. Any desired effect should be weighed against unwanted effects for the individual or for society. Evaluation should be conducted in close collaboration with local health authorities.

Apps have emerged as a new public health tool in the COVID-19 epidemic, and will be with us for a while. Digital contact tracing is a sustainable measure that can reduce levels of infection in the community and help to control COVID-19 epidemics. A rigorous assessment of its effectiveness will enable us to include app-based contact tracing and information tools into plans to prepare for future outbreaks of other infectious diseases.